\pdfoutput=1 
\documentclass[12pt]{article}

\usepackage{graphicx}
\usepackage{amsmath}
\usepackage{amssymb}
\usepackage{lineno}
\usepackage{url}
\usepackage{xspace,multicol}
\usepackage{siunitx}
\usepackage{subcaption}
\usepackage{color, colortbl}
\usepackage{ragged2e}
\usepackage{array}
\usepackage{tabularx}
\usepackage{authblk}
\usepackage{heppennames}
\usepackage{libertine}
\newcommand{\guineapig}{GuineaPig\xspace}
\newcommand{\sid}{SiD\xspace}
\newcommand{\electron}{e\textsuperscript{-}\xspace}
\newcommand{\positron}{e\textsuperscript{+}\xspace}

\newcommand{\murm}{%
  \ifmmode
    \mathchoice
        {\hbox{\normalsize\textmu}}
        {\hbox{\normalsize\textmu}}
        {\hbox{\scriptsize\textmu}}
        {\hbox{\tiny\textmu}}%
  \else
    \textmu
  \fi
}

\usepackage{fixltx2e}
\definecolor{Gray}{gray}{0.9}
\DeclareSIUnit\gauss{G}

\begin{document}


\title{Impact of the new ILC250 beam parameter set\\on the \sid vertex detector occupancy arising from \positron\electron pair background\vspace*{0.3cm}\\{\normalsize Talk presented at the\\\large International Workshop on Future Linear Colliders (LCWS2017), Strasbourg, France, 23-27 October 2017. C17-10-23.2.}}

\author[1,2]{Anne Sch\"utz}

\affil[1]{\normalsize Karlsruhe Institute of Technology (KIT), Department of Physics, Institute of Experimental Particle Physics (ETP), Wolfgang-Gaede-Str. 1, 76131 Karlsruhe}
\affil[2]{\normalsize Deutsches Elektronen-Synchrotron (DESY), Notkestr. 85, 22607 Hamburg}

\maketitle


\begin{abstract}
The Silicon Detector (SiD) is one of the detector concepts for the International Linear Collider (ILC), which is proposed to be built in Japan.
Since the decision to reduce the center-of-mass energy for the first stage of the ILC to \SI{250}{\GeV}, numerous efforts are ongoing to evaluate the possible physics program and the change in the detector background.
It was found that changing the beam parameters would increase the luminosity and therefore further strengthen the physics outcome.
On the downside, the new beam parameters lead to an increase in the detector background arising from beam-beam interactions.
\\
The simulation study, which is the topic of this note, focuses on the impact the increased background levels have on the SiD vertex detector performance.
The results suggest that the vertex detector occupancies can be accommodated in the design of the SiD detector without any loss in performance.
\end{abstract}

\section{Introduction}
\label{sec:introduction}
In November 2017, the panel of the International Committee for Future Accelerators (ICFA) made an official statement that the first stage of the International Linear Collider (ILC) will have a center-of-mass energy of \SI{250}{\GeV} instead of \SI{500}{\GeV}, with the goal to reduce the total costs of the ILC project~\cite{ICFA_Statement}.
In order to increase the luminosity of this ILC250 stage, new beam parameters were suggested at the Americas Workshop on Linear Colliders (AWLC2017) at SLAC in June 2017~\cite{AWLC_Yokoya, AWLC_Jeans}.
The changes to the parameter sets were officially approved as the Change Request CR-0016 by the Technical Change and Management Board (TCMB) in September 2017~\cite{CR-0016}.\\
The changes include the reduction of the horizontal beam emittance, which causes a rise in the luminosity.
The luminosity is the topic of the first section of this note.
The reduced beam emittance on the other hand also leads to a larger electron positron pair background from increased beam-beam interactions, and to an effect on the spatial distribution of these pairs.
The magnetic solenoid field of the detectors forces the pair background particles into helix tracks which extend to the inner layers of the detector.
Since the detector concepts proposed for the ILC will be designed to deliver measurements with unprecedented precisions, the detector geometry and the readout designs have to be carefully optimized.\\
This note presents a study of the density of the \Pep\Pem pair particles for four different beam parameter sets of the ILC250 stage.
Additionally, the vertex detector occupancy of the Silicon Detector (SiD) was analyzed in order to see the effect of the increased background on the detector.  
The performance of the vertex detector plays a crucial role for the precision of physics studies.
Therefore, the aim is to keep the overall ratio of so-called dead cells below a per mill level.
By definition, a detector pixel cell is ``dead'' when no further hits can be stored in its available buffers.\\
The plots in the sections regarding the background density and the arising detector occupancy display the envelopes of the pair helix tracks, as well as the vertex detector occupancy for the new ILC250 parameter set in comparison to other possible beam parameters.
Adapting the naming scheme of the change request CR-0016, the beam parameter schemes are as listed in Table~\ref{tab:Parameters}.
The new official beam parameters are given by set (A), for which the horizontal beam emittance  $\epsilon_x$ is reduced by half compared to the original parameters given in the Technical Design Report (TDR)~\cite[p. 11]{TDR1}.
\begin{table}[h]
\centering
\begin{tabularx}{0.52\textwidth}{llll}
\hline\hline
\textbf{Set}  & \textbf{$\epsilon_x$ [\murm m]} & \textbf{$\beta^*_x$ [mm]} & \textbf{$\beta^*_y$ [mm]}\\
\hline
\cline{1-4}
\hline
 TDR & 10 & 13.0 & 0.41\\
 \rowcolor{Gray}
 (A) & 5 & 13.0 & 0.41\\
 (B) & 5 & 9.19 & 0.41\\
 (C) & 5 & 9.19 & 0.58\\
\hline\hline
\end{tabularx}
\caption{\textit{The new official beam parameter set is set (A)~\cite{CR-0016}.
Sets (B) and (C) were further possibilities that were under discussion.
The table only lists the parameters that are to be changed with respect to the original ILC250 parameters (set (TDR)) given in the Technical Design Report (TDR)~\cite[p. 11]{TDR1}.\\
For all new sets, the horizontal emittance $\epsilon_x$ is reduced. 
Additionally, the horizontal and vertical beta functions at the IP, $\beta^*_x$ and $\beta^*_y$, are reduced for sets (B) and (C).}}
\label{tab:Parameters}
\end{table}

\newpage
\section{The luminosity spectra of the ILC250 beam parameter sets}
\label{sec:Envelopes}
\begin{figure}[h]
\centering
\begin{subfigure}[t]{0.49\textwidth}
\centering
\includegraphics[width=\textwidth] {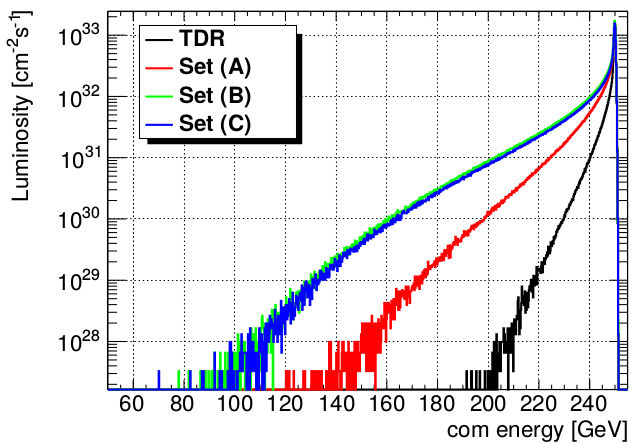}
\caption{}
\end{subfigure}
\hspace*{0.08cm}
\begin{subfigure}[t]{0.49\textwidth}
\centering
\includegraphics[width=1.02\textwidth] {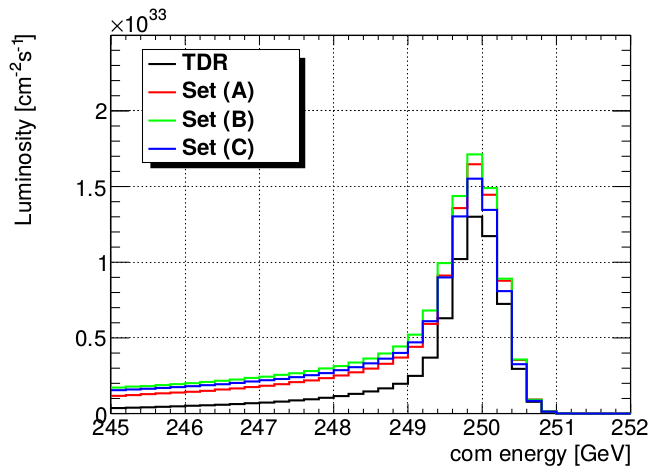}
\caption{}
\end{subfigure}
\caption{\textit{Luminosity spectra generated with \guineapig, using the ILC250 parameter sets listed in Table~\ref{tab:Parameters}.
Figure (b) shows the spectra zoomed in at around \SI{250}{\GeV}.}}
\label{fig:Lumi_Spectra}
\end{figure}

The processes, in which the beamstrahlung from beam-beam interactions produces secondary \Pep \Pem pairs, were simulated with \guineapig~\cite{Schulte:1997nga}.
\guineapig is a Monte Carlo (MC) background event generator comparable to CAIN~\cite{CAIN}, which was used for the ILC250 background studies presented at AWLC17~\cite{AWLC_Jeans}.\\
Using the beam particle distributions generated by \guineapig, the luminosity spectra of the ILC250 stage parameters can be plotted as shown in Figure~\ref{fig:Lumi_Spectra}.
\\Comparing the total luminosity resulting from the spectra in Figure~\ref{fig:Lumi_Spectra}, the expected rise in the luminosity due to the reduction of the horizontal beam emittance and the beta function is demonstrated.
\\The value for the TDR set (\SI{8.26e33}{\per\centi\meter\squared\per\second}) confirms the calculations of the luminosity done for the Change Request CR-005, which states a value of \SI{8.2e33}{\per\centi\meter\squared\per\second}~\cite{CR-005}.
The new official beam parameters of set (A) produce a luminosity of \SI{1.62e34}{\per\centi\meter\squared\per\second}, which is an increase of about \SI{97}{\percent}.
\begin{table}[h]
\centering
\begin{tabularx}{0.4\textwidth}{cc}
\hline\hline
\textbf{Set}  & Luminosity [\si{\per\centi\meter\squared\per\second}]\\
\hline
\cline{1-2}
\hline
 TDR & \num{8.26e33}\\
 \rowcolor{Gray}
 (A) & \num{1.62e34}\\
 (B) & \num{2.38e34}\\
 (C) & \num{2.15e34}\\
\hline\hline
\end{tabularx}
\caption{\textit{The total luminosities resulting from the luminosity spectra generated with \guineapig for the different ILC250 beam parameter sets.}}
\label{tab:Luminosity}
\end{table}
\section{The pair background envelopes}
\label{sec:Envelopes}
For all ILC250 parameter sets listed in Table~\ref{tab:Parameters}, a full bunch train (1312 bunch crossings) of the pair background particles was generated.
The \guineapig generation does not take the crossing angle of \SI{14}{\milli\radian} into account.
The calculation of the helix tracks that the pairs form in the SiD solenoid field with a strength of \SI{5}{\tesla} was done using the algorithm described in a former note presenting the pair envelopes of the ILC500 scheme~\cite{HelixProceeding}.

\begin{figure}
\centering
\begin{subfigure}[t]{0.49\textwidth}
\centering
\includegraphics[width=\textwidth] {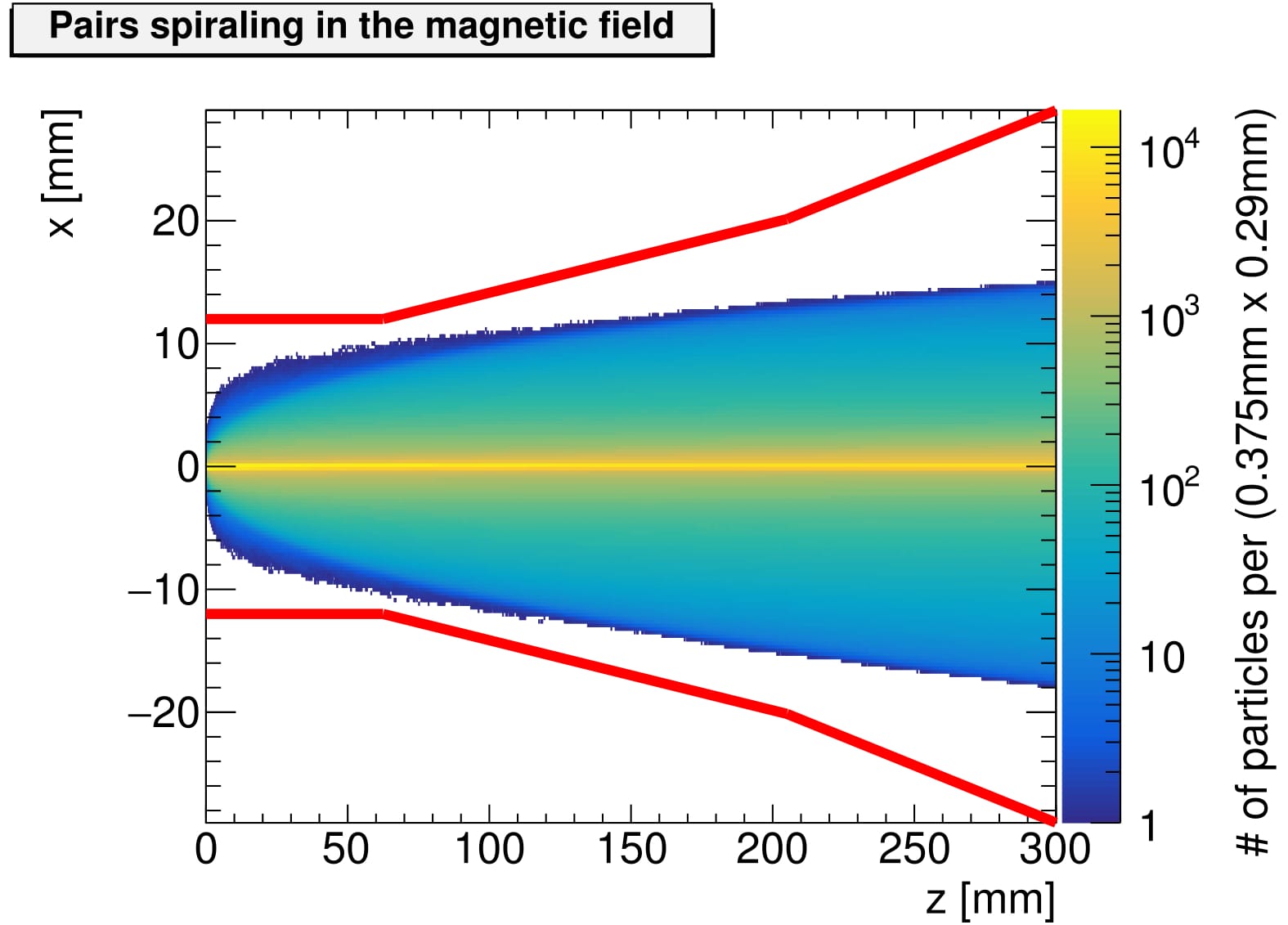}
\caption{\textit{ILC250 set (TDR)}}
\end{subfigure}
\hspace*{0.08cm}
\begin{subfigure}[t]{0.49\textwidth}
\centering
\includegraphics[width=\textwidth] {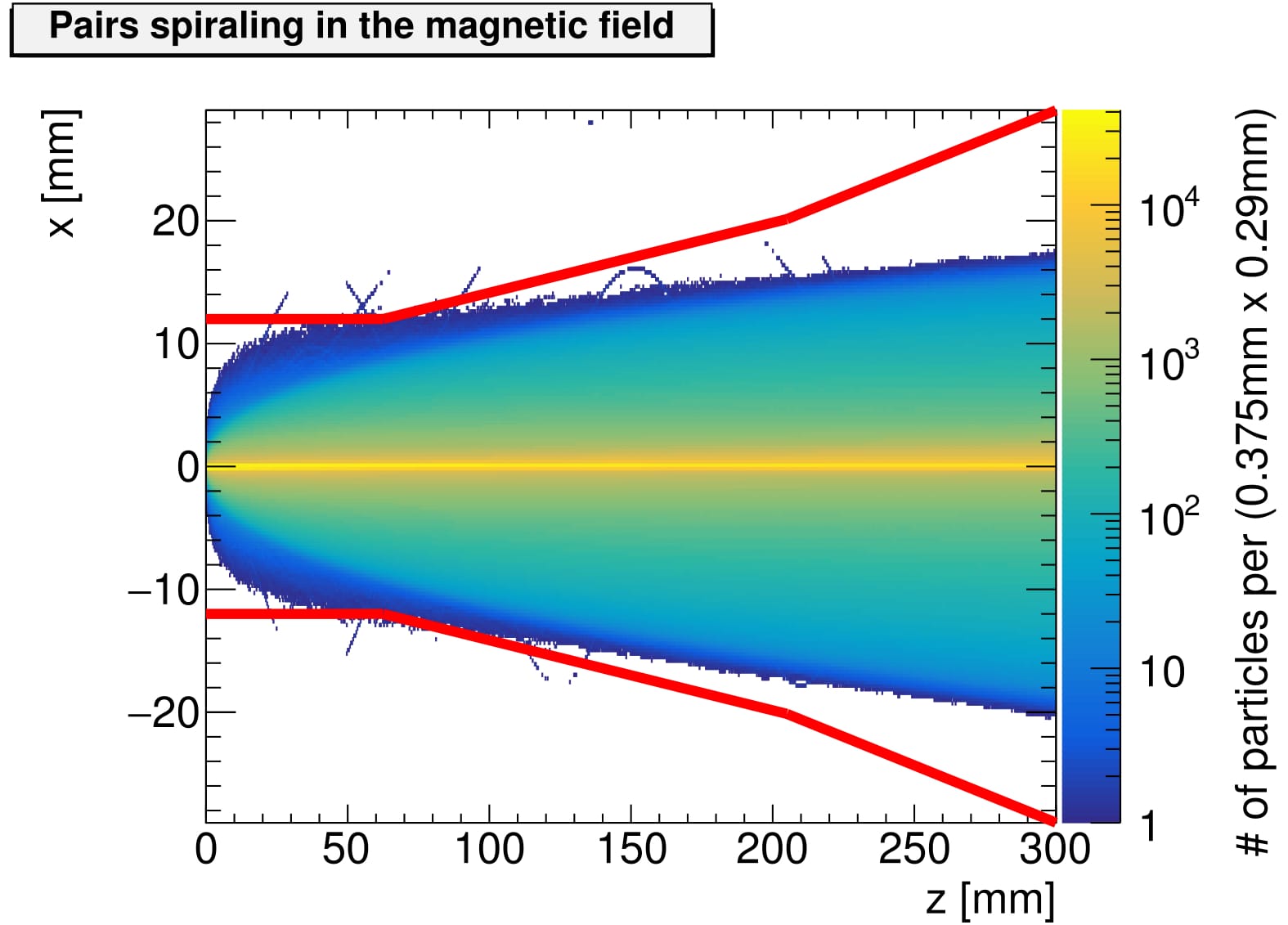}
\caption{\textit{ILC250 set (A)}}
\end{subfigure}
\\
\begin{subfigure}[t]{0.49\textwidth}
\centering
\includegraphics[width=\textwidth] {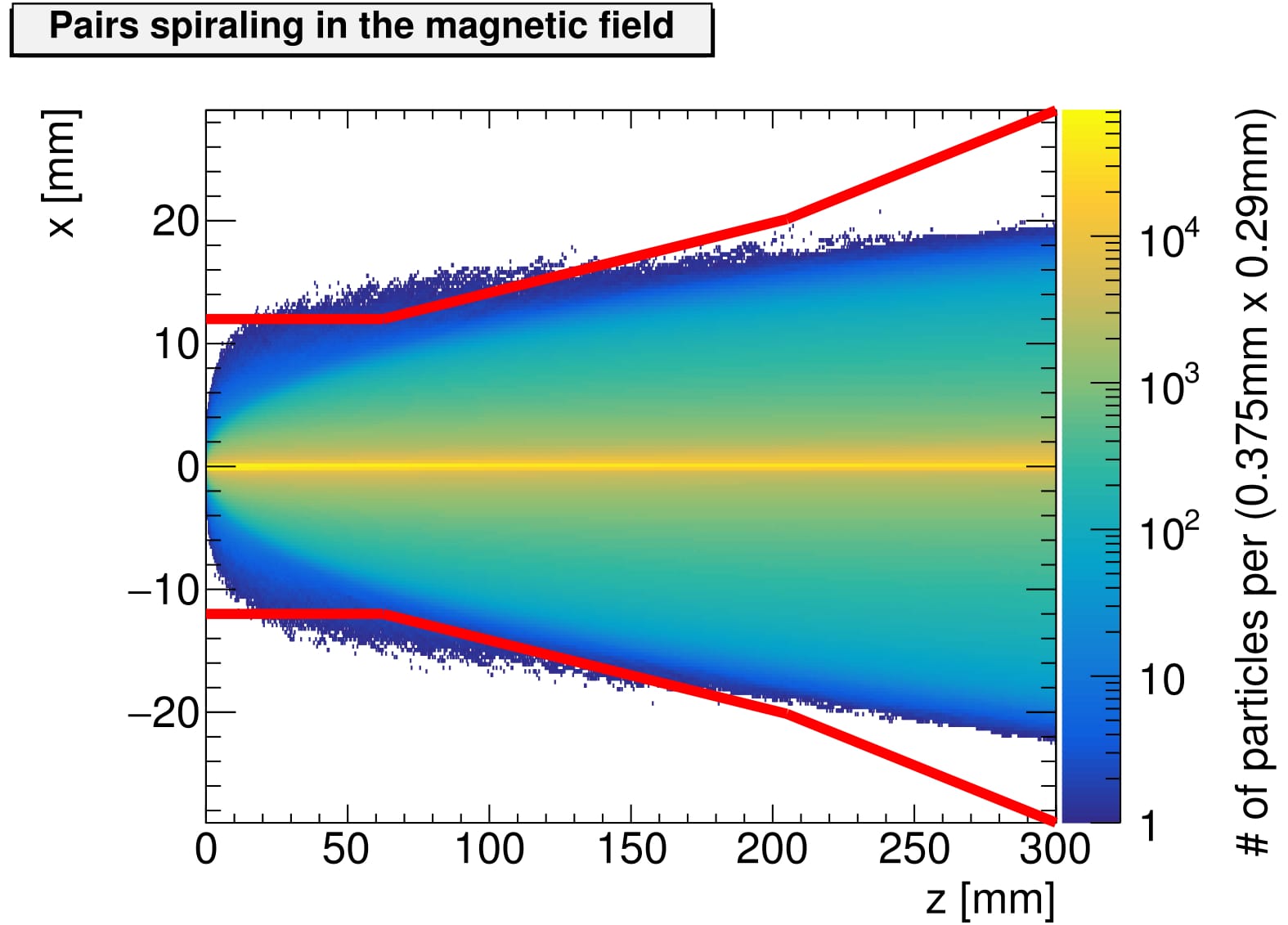}
\caption{\textit{ILC250 set (B)}}
\end{subfigure}
\hspace*{0.08cm}
\begin{subfigure}[t]{0.49\textwidth}
\centering
\includegraphics[width=\textwidth] {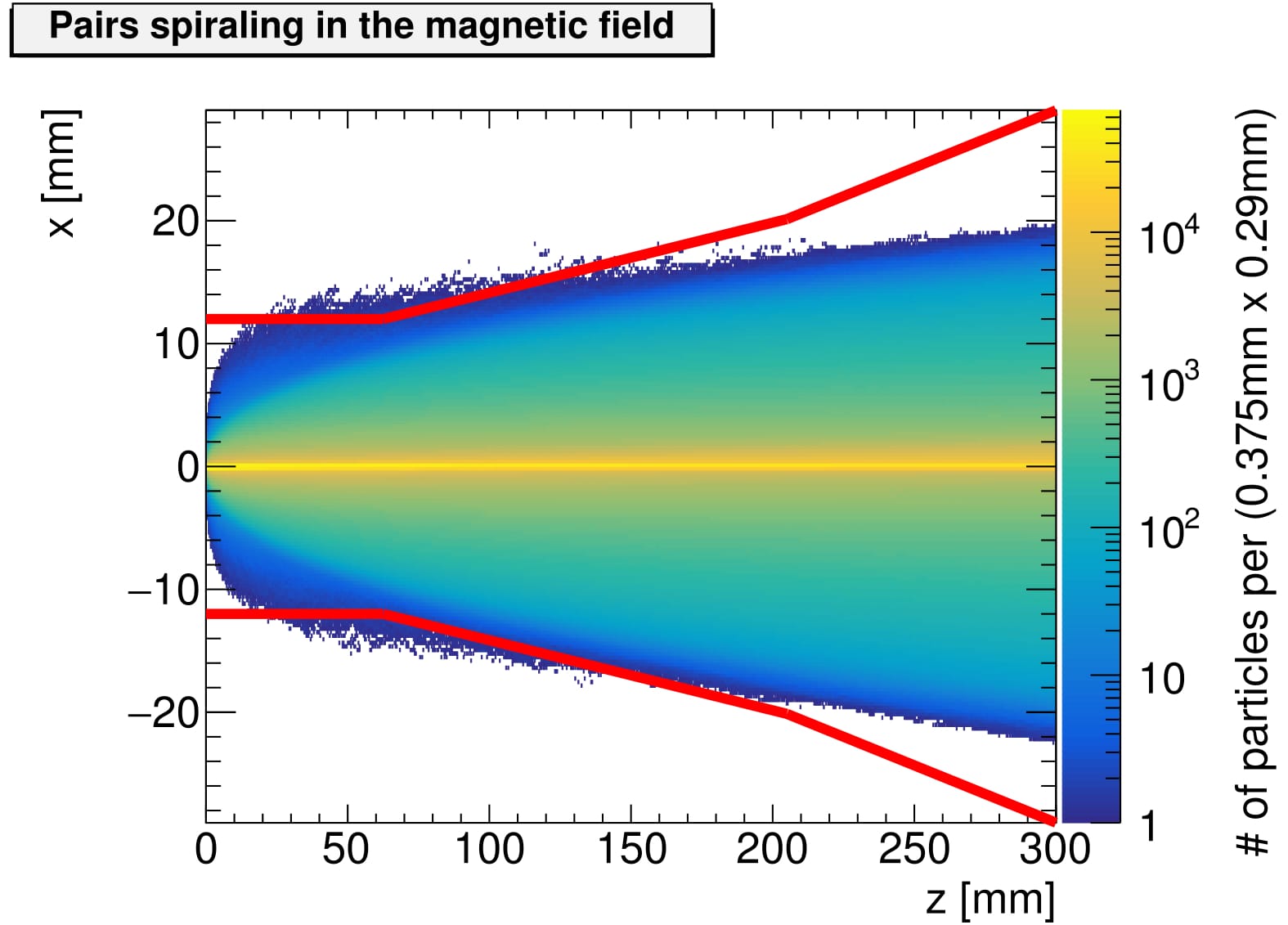}
\caption{\textit{ILC250 set (C)}}
\end{subfigure}
\caption{\textit{Pair background density for the different ILC250 beam parameter sets for a full bunch train (1312 bunch crossings). Although the pairs are deflected in the SiD solenoid field, the high-p\textsubscript{T} pairs reach past the beam pipe towards the inner layers of the detector. The beam pipe is represented by the red solid lines.}}
\label{fig:Envelopes}
\end{figure}

\begin{figure}
\centering
\includegraphics[width=0.78\textwidth] {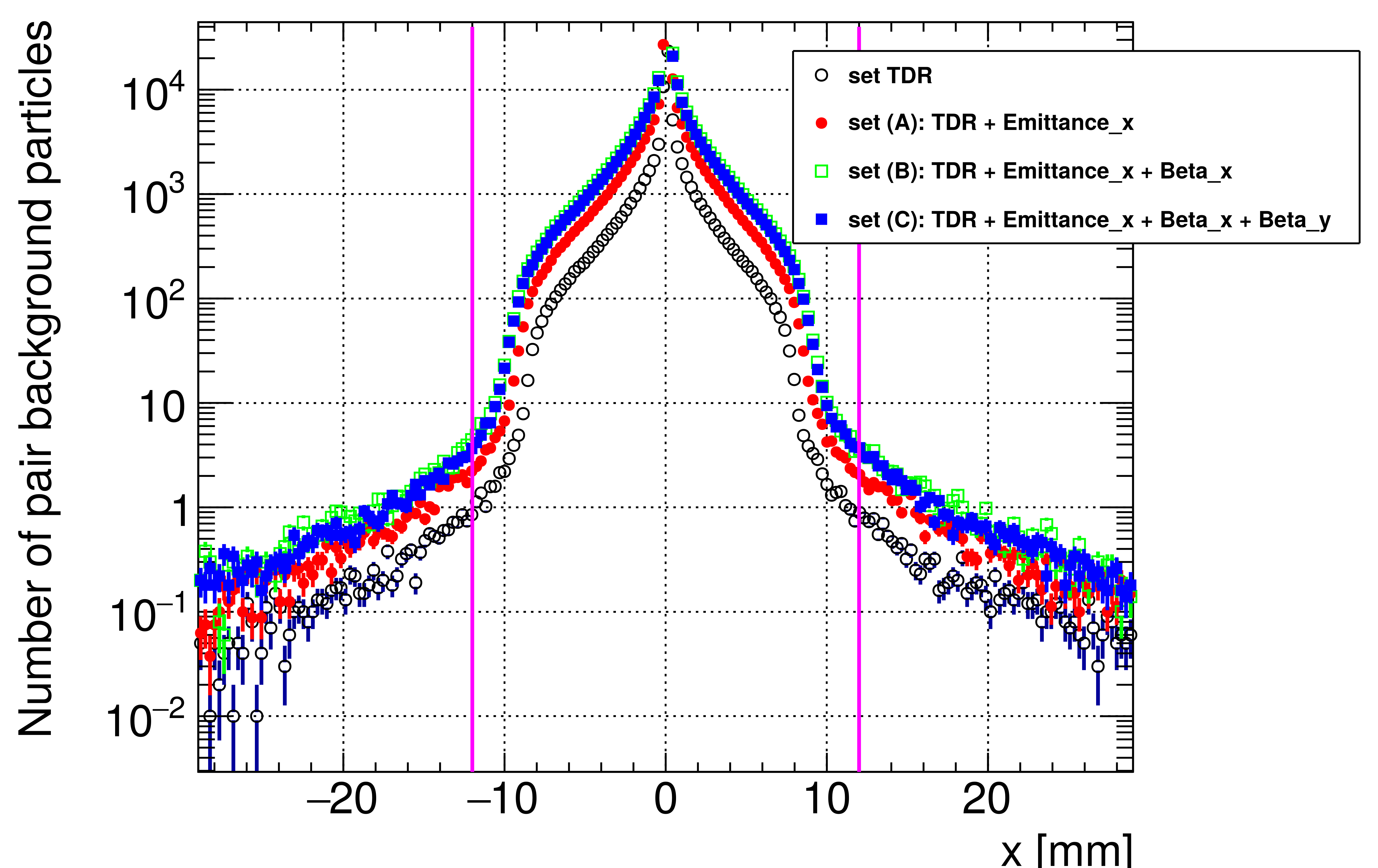}
\caption{\textit{Projection of the pair background density for a full bunch train (1312 bunch crossings), which is shown in Figure~\ref{fig:Envelopes}, along the x direction.
The projection is done for z $=$ \SI{62}{\milli\meter}, the z position of the first beam pipe kink.
At that position, the background envelopes are the largest compared to the beam pipe radius.
The beam pipe is indicated by the pink solid lines.}}
\label{fig:Projection_Envelopes}
\end{figure}

Figure~\ref{fig:Envelopes} shows the density of pairs in the area around the interaction point.
Since the particles are deflected in the solenoid field, they form helix tracks that reach out towards the innermost layers of the detector.
The red solid lines in the plots represent the outline of the beam pipe.
The beam envelopes of schemes (B) and (C) are significantly expanded with respect to the beam envelopes of the TDR scheme, whilst the pair envelope for set (A) is still contained within the beam pipe.
The projection of the pair background density along $x$ (Figure~\ref{fig:Projection_Envelopes}) shows that the number of pairs in set (A) is increased by a factor of 2-3 in comparison to the TDR scheme, and by a factor of 6-7 in sets (B) and (C).
Nevertheless, the envelopes for all parameter schemes are well contained within the beam pipe.
At a beam pipe radius of \SI{12}{\milli\meter}, one can expect less than 10 interactions of pairs with the material of the beam pipe per bunch crossing.

\section{SiD vertex detector occupancies}
\begin{figure}
\centering
\begin{subfigure}[t]{0.45\textwidth}
\centering
\includegraphics[width=1.05\textwidth]{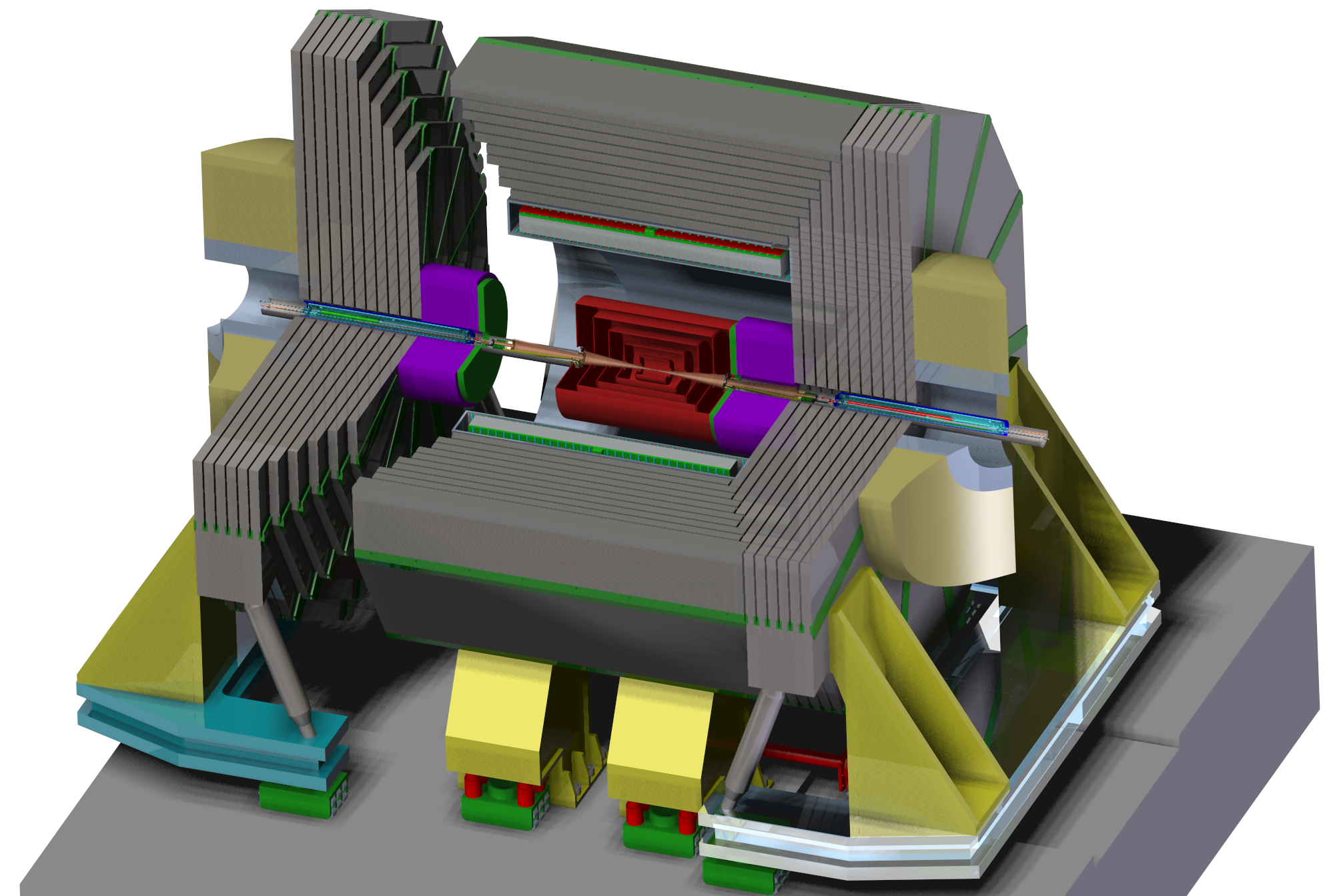}
\caption{\textit{Visualization of the full SiD detector}}
\end{subfigure}
\hspace*{0.3cm}
\begin{subfigure}[t]{0.45\textwidth}
\centering
\includegraphics[width=\textwidth]{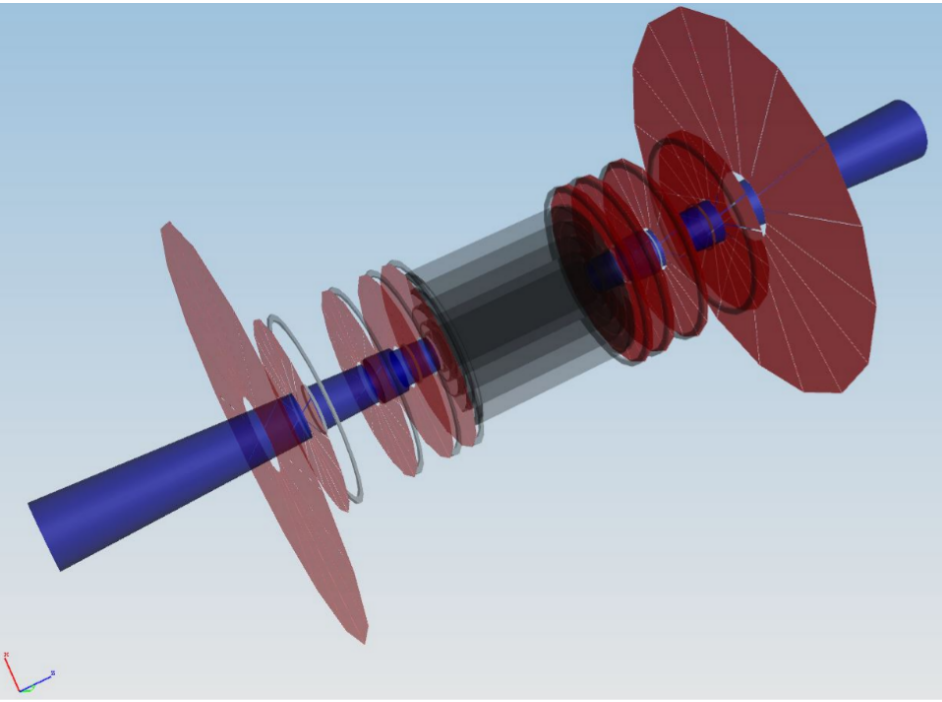}
\caption{\textit{Visualization of the SiD vertex detector}}
\end{subfigure}
\caption{\textit{Visualizations of the current design models of the SiD detector (a), and the SiD vertex detector (b).}}
\label{fig:SiD}
\end{figure}

The occupancy of the SiD vertex detector is calculated by counting the hits from the pair background per pixel for a full bunch train (1312 bunch crossings).
The sensor buffers are read out after every train, and therefore have to be suitable for the expected detector occupancy.
For the study of the occupancy, the pair background events were taken from GuineaPig, as described in Section \ref{sec:Envelopes}, and used as input for a full Geant4~\cite{geant_ref,geant_ref2} detector simulation with SLIC~\cite{SLIC}.
The crossing angle was applied in this simulation step.\\
Regarding the SiD detector geometry, a pixel size of \SI{20x20}{\micro\meter} was assumed for the vertex detector.
The SiD vertex detector is a high-resolution pixel detector with the following requirements:
\begin{itemize}
 \item Spatial hit resolution: $<$ \SI{4}{\micro\meter}
 \item Momentum resolution: $\sim$ 2-\SI{5e-5}{\GeV}
 \item $\sim$ \SI{0.1}{\percent} X\textsubscript{0} per layer
\end{itemize}
Table~\ref{tab:KeyParametersSiD} lists further key parameters and measurements of the vertex detector, for which a visualization is shown in Figure~\ref{fig:SiD}.
\begin{table}[h]
\centering
\begin{tabularx}{\textwidth}{l|llll}
\hline\hline
& Technology & Inner radius & Outer radius & z extent\\
\hline
VXD barrel & Silicon pixels & 1.4 & 6.0 & $\pm 6.25$ \\
\hline\hline
& Technology & Inner z & Outer z & Outer radius\\
\hline
VXD endcap & Silicon pixels & 7.3 & 83.4 & 16.6 \\
\hline\hline
\end{tabularx}
\caption{\textit{Key design parameters of the baseline SiD vertex detector (VXD). All dimensions are given in cm.}}
\label{tab:KeyParametersSiD}
\end{table}

\newpage
\subsection{Occupancy studies for the current SiD design}
\begin{figure}[h]
\centering
\begin{subfigure}[t]{0.45\textwidth}
\centering
\includegraphics[width=1.06\textwidth]{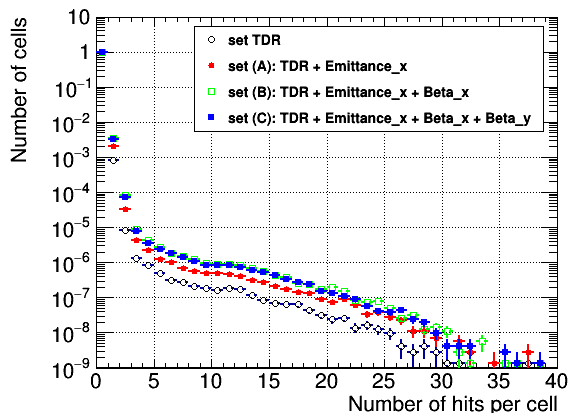}
\caption{\textit{Normalized Occupancy:\\Number of cells containing a certain amount of hits, normalized by the total number of cells of the vertex detector.}}
\end{subfigure}
\hspace*{0.3cm}
\begin{subfigure}[t]{0.45\textwidth}
\centering
\includegraphics[width=1.05\textwidth]{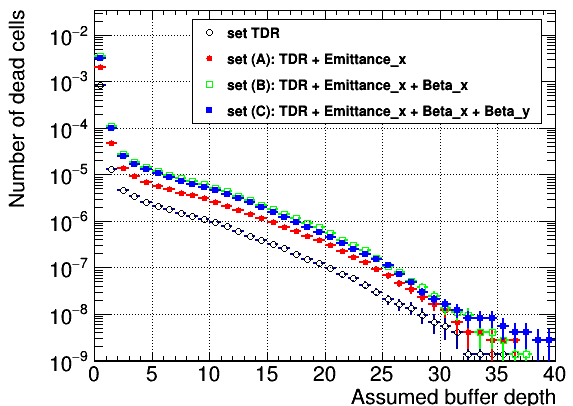}
\caption{\textit{Ratio of dead cells in comparison to all cells of the vertex detector.}}
\end{subfigure}
\caption{\textit{Pair background occupancy in the SiD vertex detector for all vertex detector layers combined, after a full bunch train (1312 bunch crossings).}}
\label{fig:All_layers_Occupancy}
\end{figure}

\begin{figure}[h!]
\centering
\begin{subfigure}[t]{0.45\textwidth}
\centering
\includegraphics[width=1.05\textwidth]{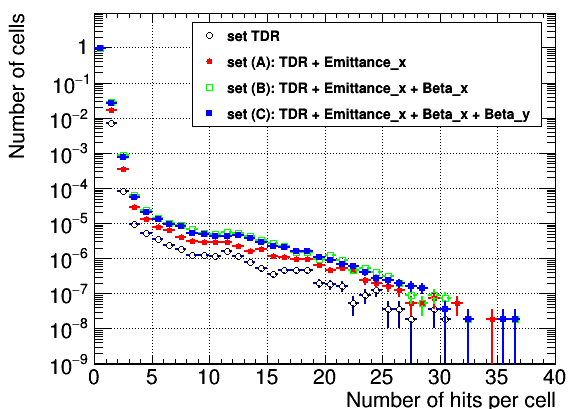}
\caption{\textit{Layer 0: Normalized Occupancy}}
\end{subfigure}
\hspace*{0.3cm}
\begin{subfigure}[t]{0.45\textwidth}
\centering
\includegraphics[width=1.05\textwidth]{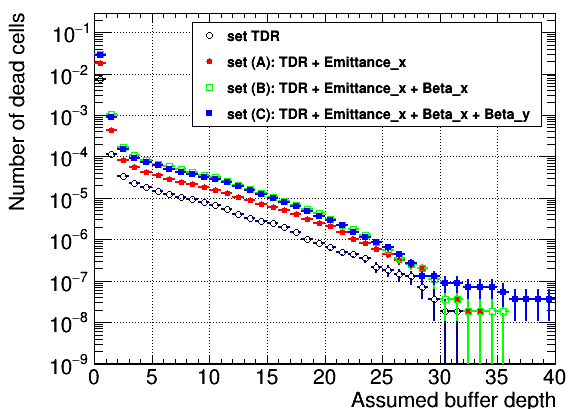}
\caption{\textit{Layer 0: Ratio of dead cells}}
\end{subfigure}
\caption{\textit{Pair background occupancy in the SiD vertex detector layer 0, after a full bunch train (1312 bunch crossings).
The plot on the left hand side shows the normalized occupancy, the ratio of dead cells is shown on the right side.}}
\label{fig:Layer0_Occupancy}
\end{figure}

The current SiD design with an L$^*$ of \SI{4.1}{\meter} (the distance between the interaction point and the last quadrupole magnet of the Final-Focus system) foresees an anti-DiD magnet with a magnetic field strength of \SI{600}{\gauss}.
It reduces the pair background occupancy by sweeping the pair background particles through the outgoing beam pipe.
The following plots show occupancy studies using the current SiD geometry.\\
On the left hand side of Figure~\ref{fig:All_layers_Occupancy}, the normalized occupancy for the full vertex detector is shown.
From that, the ratio of dead cells was calculated (right).
A cell is ``dead'' when the number of hits in this cell has reached the cell's buffer depth.
Therefore, the number of dead cells is plotted as a function of the assumed buffer depth for the vertex detector cells.
For all layers combined, the new parameter scheme (A) leads to a fraction of dead cells that is 2-3 times higher than for the TDR set.
Again, for the schemes (B) and (C), the fraction is even higher.
\\In general, the occupancy in layer 0 is higher than in the other layers, as it is the layer closest to the interaction point.
Here, the occupancy for the new sets is significantly increased with respect to the TDR scheme (see Figure~\ref{fig:Layer0_Occupancy}). 

\subsection{Occupancy studies as a comparison of different SiD designs}
\begin{figure}[!h]
\centering
\begin{subfigure}[t]{0.45\textwidth}
\centering
\includegraphics[width=1.05\textwidth]{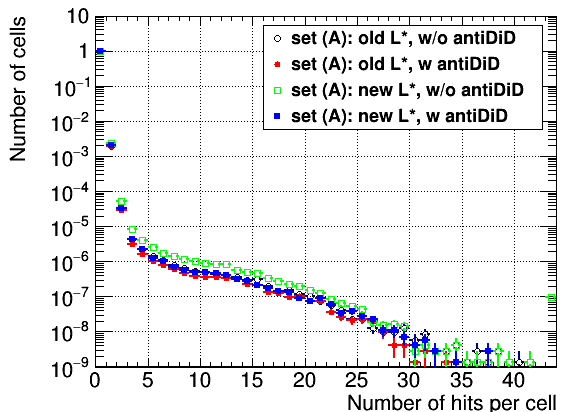}
\caption{\textit{All layers combined}}
\end{subfigure}
\hspace*{0.3cm}
\begin{subfigure}[t]{0.45\textwidth}
\centering
\includegraphics[width=1.05\textwidth]{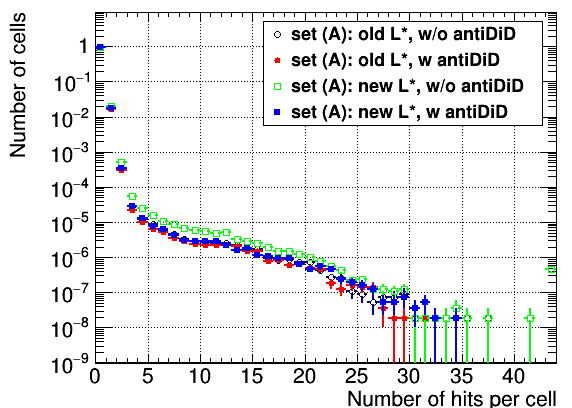}
\caption{\textit{Layer 0}}
\end{subfigure}
\caption{\textit{Pair background occupancy in the SiD vertex detector for different SiD designs, after a full bunch train (1312 bunch crossings).
The plot on the left hand side shows the normalized occupancy for all layers combined, whilst the normalized occupancy for only layer 0 is shown on the right.}}
\label{fig:SiD_comparison}
\end{figure}

In the following, different SiD designs are compared by observing the vertex detector occupancy for the new ILC250 parameter set (A).
The studied designs are different combinations of the old L$^*$ value (\SI{3.5}{\meter}) vs. the new value, and of the inclusion or exclusion of the anti-DiD field.
Again, the normalized occupancy is plotted for all vertex detector layers combined, as well as only for layer 0 (see Figure~\ref{fig:SiD_comparison}).
\\Figure~\ref{fig:SiD_comparison} shows that the SiD design does not have a big impact on the vertex detector occupancy from the pair background. 
The lowest occupancy is reached for the SiD design with the old L$^*$ and anti-DiD.\\
However, the current SiD design (new L$^*$ with anti-DiD) shows occupancy levels that are only marginally worse.
Assuming a buffer depth of four, just above 10$^{-5}$ of all cells in layer 0 would have full buffers, filled with exactly four hits.
With the rule of thumb that the vertex detector occupancy for backgrounds should be below 10$^{-4}$, the occupancies for all studies shown in this note are well below this boundary.
\section{Summary and conclusion}
By looking at the luminosity spectra generated for this study, the expected rise in luminosity for the proposed ILC250 parameter sets is confirmed.
The total luminosity resulting from the new official beam parameter set (A) is increased by about \SI{97}{\percent} with respect to the original beam parameters, leading to a new value of \SI{1.62e34}{\per\centi\meter\squared\per\second}.
The background studies indicate that the new beam parameters increase the \Pep\Pem pair background density and the SiD vertex detector occupancy by only a factor of about 2-3 in comparison to the original TDR beam parameters.
As for a precision detector the aim is to keep the overall ratio of so-called ``dead'' cells with a full buffer below a per mill level, any rise in the occupancy level can be critical.
Although the density of the pair background close to the interaction point rises, the normalized vertex detector occupancy for the new set is well below 10$^{-4}$.
The highest occupancy is reached in the innermost layer with a fraction of dead cells of about \num{5e-5}, assuming a buffer depth of four.\\
The SiD Optimization group is confident that this rise in the occupancy can be accommodated in the design of the SiD vertex detector without a loss in precision for the physics studies.
Therefore, the SiD group welcomes the decision to change the beam parameters in order to gain higher luminosities and to strengthen the physics program.

\section*{Acknowledgments}
The author would like to thank the SiD Optimization group, as well as Daniel Jeans (KEK) for useful discussions and helpful input to the presented studies.

\bibliographystyle{unsrt}
\bibliography{bibliography.bib}

\begin{thebibliography}{10}

\bibitem{ICFA_Statement}
{ICFA panel}.
\newblock {ICFA Statement on the ILC Operating at 250 GeV as a Higgs Boson
  Factory}.
\newblock LCB Conclusion Report,
  \url{http://icfa.fnal.gov/wp-content/uploads/ICFA-Statement-Nov2017.pdf},
  November 2017.

\bibitem{AWLC_Yokoya}
Kaoru Yokoya.
\newblock {Overview of ILC optimization at the center of mass energy of
  250GeV}.
\newblock
  \url{https://agenda.linearcollider.org/event/7507/contributions/39290/attachments/31821/47993/HigherLumi250GeV-AWLC2017-Yokoya.pdf},
  June 2017.
\newblock {Contribution to Americas Workshop on Linear Colliders AWLC17, June
  26-30, 2017 SLAC, Menlo Park, CA}.

\bibitem{AWLC_Jeans}
Daniel Jeans.
\newblock {Physics impact of 250 GeV Optimization}.
\newblock
  \url{https://agenda.linearcollider.org/event/7507/contributions/39291/attachments/31828/48012/mdi-beamparams-awlc-2017.pdf},
  June 2017.
\newblock {Contribution to Americas Workshop on Linear Colliders AWLC17, June
  26-30, 2017 SLAC, Menlo Park, CA}.

\bibitem{CR-0016}
ILC Technical and Change~Management Board.
\newblock {Change Request No. ILC-CR-0016, Luminosity Improvement at 250GeV
  CM}.
\newblock EDMS No: D*1159725, September 2017.

\bibitem{TDR1}
{Behnke, T. et al. [ILC Collaboration]}.
\newblock {\em {The International Linear Collider Technical Design Report -
  Volume 1: Executive Summary}}.
\newblock 2013.
\newblock arXiv:1306.6327 [physics.acc-ph].

\bibitem{Schulte:1997nga}
Daniel Schulte.
\newblock {\em {Study of Electromagnetic and Hadronic Background in the
  Interaction Region of the TESLA Collider}}.
\newblock PhD thesis, {DESY}, 1997.

\bibitem{CAIN}
Kaoru Yokoya.
\newblock {User's Manual of CAIN, Version 2.42}.
\newblock \url{https://ilc.kek.jp/~yokoya/CAIN/Cain242/CainMan242.pdf}, June
  2011.

\bibitem{CR-005}
ILC Technical and Change~Management Board.
\newblock {Change Request No. ILC-CR-005, Update of published ILC top-level
  luminosity parameters}.
\newblock EDMS No: D*01100895, April 2015.

\bibitem{HelixProceeding}
{Anne Sch\"utz}.
\newblock {Pair Background Envelopes in the SiD Detector}.
\newblock Proceedings of the International Workshop on Future Linear Colliders
  (LCWS2016), Morioka, Japan, 5-9 December 2016. C16-12-05.4, 2017.
\newblock arXiv:1703.05737.

\bibitem{geant_ref}
S.~Agostinell et~al.
\newblock {GEANT4: A Simulation toolkit}.
\newblock {\em Nucl. Instrum. Meth. A}, 506:250--303, 2003.

\bibitem{geant_ref2}
J.~Allison et~al.
\newblock {Geant4 developments and applications}.
\newblock {\em {IEEE Transactions on Nuclear Science}}, 53(1):270--278, Feb
  2006.

\bibitem{SLIC}
N.~Graf and J.~McCormick.
\newblock {Simulator for the linear collider (SLIC): A tool for ILC detector
  simulations}.
\newblock {\em AIP Conf. Proc.}, 867:503--512, 2006.
\newblock [,503(2006)].

\end{thebibliography}

\end{document}